\documentclass[aps,prl,twocolumn]{revtex4}
\usepackage{amsfonts}
\usepackage{amsmath}
\usepackage{amssymb}
\usepackage{graphicx}

\begin{document}

\title{Origin of Low Thermal Conductivity in Nuclear Fuels}
\author{Quan Yin and Sergey Y. Savrasov}
\affiliation{Department of Physics, University of California, Davis, CA 95616}

\begin{abstract}
Using a novel many-body approach, we report lattice dynamical properties of
UO$_{2}$ and PuO$_{2}$ and uncover various contributions to their thermal
conductivities. Via calculated Gr\"{u}neisen constants, we show that only
longitudinal acoustic modes having large phonon group velocities are
efficient heat carriers. Despite the fact that some optical modes also show
their velocities which are extremely large, they do not participate in the
heat transfer due to their unusual anharmonicity. Ways to improve thermal
conductivity in these materials are discussed.
\end{abstract}

\date{\today}
\maketitle

Today's nuclear fuels are based on $^{235}$U and $^{239}$Pu elements where
in a typical set-up, a nuclear reaction heats up a pellet made of either UO$%
_{2}$ or its mixture with PuO$_{2}$ and the heat is transformed to
electrical energy. One of the major issues is to conduct the heat from a
core of the pellet to its outer area which makes the evaluation of a high
temperature thermal conductivity a key problem. Unfortunately, the thermal
conductivity of UO$_{2}$ is very low and as a result, even a gradual
increase of its value may lead to significant breakthrough in the
performance of commercial nuclear reactors currently operating worldwide.

In the present work, we argue that it is the unexpectedly large
anharmonicity of optical modes that results in a very low thermal
conductivity of modern nuclear fuels. Consider semiconducting UO$_{2}$ which
is a main element of UOX fuel, or a blend of U and Pu oxides which is a
major substance of MOX fuel. The heat from the core of the pellet in these
insulating systems is carried by phonons which are known to be very
inefficient heat conductors. This brings a whole set of complex problems
such as causing fuel pellets to crack and degrade prematurely, necessitating
replacement before the fuel has been depleted.

Unfortunately, studying thermal conductivity \cite{1} as well as such
properties as crystal structures, phase diagrams, lattice dynamics and
structural instabilities of the actinide based materials is a formidable
theoretical problem. Most of the previous works have concentrated on
molecular dynamic simulations with empirically adjusted interatomic
potentials \cite{2}\cite{3}. However, the $5f$ electrons in actinides are
close to a localization-delocalization or Mott transition as it has been
recently demonstrated for Plutonium metal \cite{4}\cite{5}\cite{6}. UO$_{2}$
and PuO$_{2}$ are Mott-Hubbard insulators with energy gaps at both low and
high temperatures. Despite an impressive set of past theoretical studies 
\cite{7}\cite{8}\cite{9}\cite{10}\cite{11}, the spectral functions of the
high temperature paramagnetic regime cannot be obtained by static-mean field
theories such as the Density Functional Theory \cite{12} and require a
genuine many-body treatment.

In the present work we use a novel electronic structure method \cite{13}
capable of describing Mott insulating materials in order to address the
structural properties and thermal conductivity of the modern nuclear fuels.
Both UO$_{2}$ and PuO$_{2}$ are calculated using a combination of local
density approximation (LDA) \cite{12} and Dynamical Mean Field Theory (DMFT) 
\cite{14}\cite{15} where relativistic $5f$ shells of Uranium and Plutonium
atoms are treated by exact diagonalization of corresponding many-body
Hamiltonians obtained by allowing a hybridization between the $5f$-electrons
and the nearest oxygen $2p$ orbitals.

It is well known that a strong spin-orbit coupling of about $1eV$ present in
actinides splits 14-fold degenerate $f$ level onto $f_{5/2}$ and $f_{7/2}$
states. Group theoretical considerations assume that under cubic crystal
symmetry, the $f_{5/2}$ 6-fold degenerate level is further split onto $%
\Gamma_{8}$ quadruplet and $\Gamma_{7}$ doublet. In both UO$_{2}$ and PuO$%
_{2}$, the $\Gamma_{8}$ level comes approximately $0.1eV$ below the $%
\Gamma_{7}$ state, and valence arguments make it occupied by two electrons
for the case of UO$_{2}$ and fully occupied by four electrons for the case
of PuO$_{2}$. This sequence of the levels dictates the low temperature
properties of these two materials: The UO$_{2}$ becomes magnetic where the
9-fold degenerate many-body ground state of the atomic $f$-shell $^{3}$H$%
_{4} $ with $J=4$ is split onto 4 subsets, among which the lowest triplet
state of $\Gamma_{5}$ symmetry carries the moment of about $1.7-1.8\mu_{B}$
below the N\'{e}el temperature of $30.8K$ \cite{16}. On the other hand, PuO$%
_{2}$ has the filled $\Gamma_{8}$ one-electron level, making its atomic $%
^{3} $K$_{4}$ multiplet split in such way that the non-magnetic $\Gamma_{1}$
singlet comes lowest.

The cluster exact diagonalizations which are carried out in our LDA+DMFT
calculations support this atomic physics picture and only marginally alter
the self-energies for the $f$-electrons from their corresponding atomic
values. In fact, the crystal structures of both materials show that the
actinide elements are centered in the cubic environment (see Fig.1) assuming
that only relativistic $\Gamma _{7}$ orbital (its shape is shown in Fig.1)
of the $f_{5/2}$ state is properly coordinated by and strongly hybridized
with the O $2p$ states. The corresponding hybridization functions $\Delta
_{\alpha }(\omega )$\ are well fit by the single pole approximation, $\Delta
_{\alpha }(\omega )=\left\vert V_{\alpha }\right\vert ^{2}/(\omega
-P_{\alpha })$, producing only significant values of the matrix elements $%
V_{\alpha }\sim 1eV$ for $\Gamma _{7}$ in the $j=5/2$ manifold and several
times smaller values for the $\Gamma _{8}$ orbitals. Similar effect is seen
for $\Gamma _{6}$, $\Gamma _{7}$, $\Gamma _{8}$ states in the $j=7/2$
manifold. The position of the dispersive oxygen band centered at the pole $%
P_{\alpha }$ comes out $-1\div -5eV$ below the occupied f states, making
both materials classical Mott-Hubbard insulators. We find that the
fundamental energy gaps produced by the lower (occupied) and upper
(unoccupied) Hubbard bands are equal to $2.2eV$ for UO$_{2}$ and $2.5eV$ for
PuO$_{2}$. The experimental energy gap in UO$_{2}$ is of the order of $2eV$ 
\cite{16}. These data are obtained utilizing a recently developed matrix
expansion algorithm \cite{17} which helps to perform full self-consistency
with respect to both the charge densities as required by the LDA procedure
and the $5f$-electron spectral functions as required by the DMFT. Here, the $%
5f$-electron self-energies extracted from the cluster exact diagonalization
are subsequently fit using three-pole interpolation. An effective parameter $%
U_{eff}=3eV$ describing the on-site Coulomb repulsion among the $5f$
electrons is used while the other Slater integrals ($F^{(2)}$, $F^{(4)}$ and 
$F^{(6)}$) are computed from atomic physics, and are subsequently rescaled
to $80\%$ of their values to account for the effect of screening \cite{18}. 
\begin{figure}[tbp]
\begin{center}
\includegraphics[height=2.6301in,width=2.6027in]{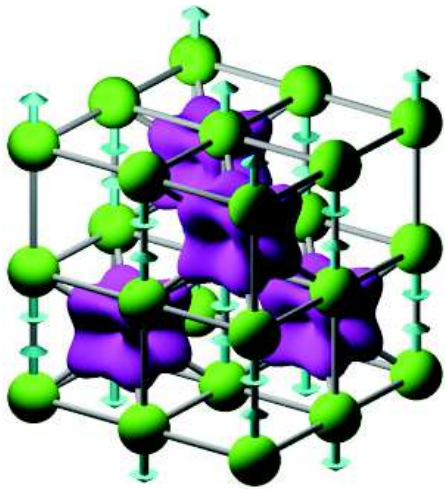}
\end{center}
\caption{Crystal structure of UO$_{2}$ and PuO$_{2}$: Uranium or Plutonium
atoms (their shapes are shown by the $\Gamma _{7}$ relativistic orbitals)
are centered inside the cubic lattice made by oxygen atoms. The arrows show
the displacements of oxygens associated with the dispersive longitudinal
optical mode.}
\label{str}
\end{figure}

In order to calculate the phonon spectra of UO$_{2}$ and PuO$_{2}$ we
utilize a new relativistic linear response phonon method which has been
recently generalized for the LDA+DMFT scheme \cite{19} and has recently
demonstrated its predicted power by finding the phonon spectrum of $\delta$%
-Pu metal \cite{20} prior to the experiment \cite{21}. In this linear
response calculation the pole interpolation for the self-energies reduces
the calculation of the dynamical matrix to standard linear response theory.
Since our interest lies in the thermal conductivity at operating
temperatures of nuclear fuels, we perform the calculations of the phonon
dispersions for paramagnetic phases of the materials with the f-electron
self-energies extracted at $T=1000K$, where not only the many-body ground
states ($\Gamma_{5}$ for UO$_{2}$ and $\Gamma_{1}$ for PuO$_{2}$) but also
various low-lying excited states are beginning to contribute.

Fig.2 shows the comparison between our calculated (filled symbols connected
by lines) and measured (open symbols) \cite{16} phonon spectrum for UO$_{2}$
along two symmetry lines of the Brillouin Zone. Fig.3 presents our predicted
phonon spectrum for PuO$_{2}$. Overall, both materials demonstrate very
similar dispersions. Due to various numerical inaccuracies and approximate
treatment of many-body effects, which, in particular, involves the use of $%
U_{eff}$, the overall mismatch between our theory and experiment for UO$_{2}$
is about $15\%$. The same accuracy can be assumed for our predicted phonon
dispersions in PuO$_{2}$. Increasing the $U_{eff}$ to $4-5$ eV marginally
affects the overall phonon spectra although makes all modes somewhat harder. 
\begin{figure}[ptb]
\begin{center}
\includegraphics[height=2.7389in,width=3.4786in]{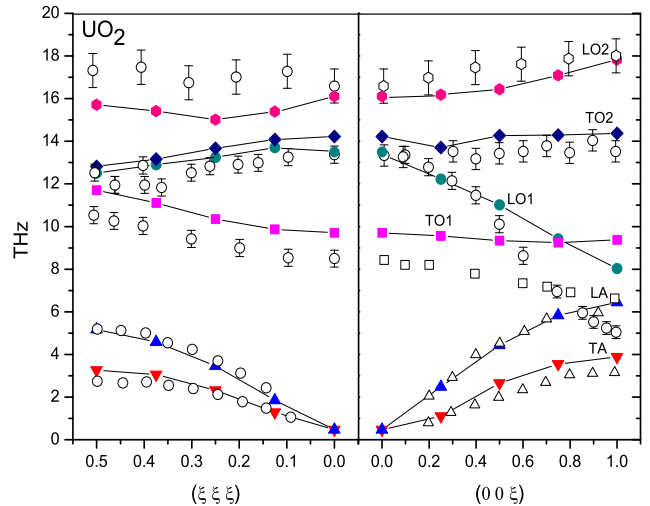}
\end{center}
\caption{Calculated phonon dispersions (filled symbols connected by lines)
and experimentally measured phonons (open symbols) \protect\cite{16} for UO$%
_{2}$.}
\label{UO2}
\end{figure}
\begin{figure}[ptb]
\begin{center}
\includegraphics[height=2.826in,width=3.4786in]{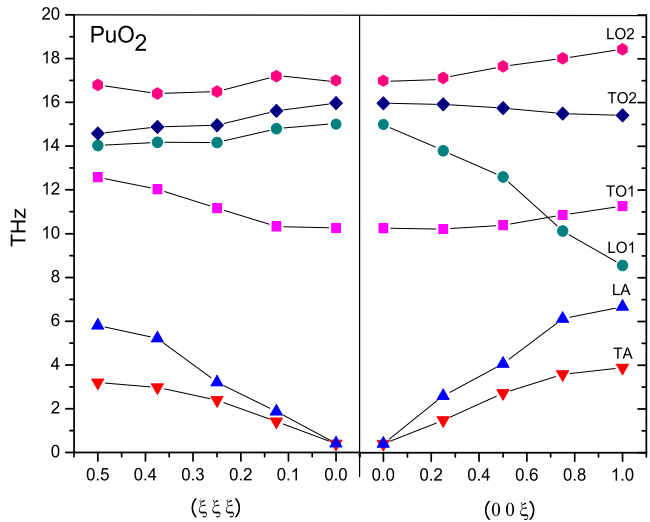}
\end{center}
\caption{Calculated phonon dispersions (filled symbols connected by lines)
for PuO$_{2}$. No experiment available.}
\label{PuO2}
\end{figure}

Our calculations reveal the following features of lattice dynamics in these
materials. First, acoustic modes are located at the low frequency region of
the spectrum producing the group velocities of about $700m/s$ for transverse
acoustic (TA) and $1100m/s$ for longitudinal acoustic (LA) modes (see also
Table 1). There are two groups of transverse and longitudinal optical modes
which we schematically label as TO1, LO1 and TO2, LO2. The first group
represents lattice vibrations associated with the out-of-phase displacements
of oxygens and little involvement of U or Pu atoms. The second group
represents the lattice vibrations of oxygens which vibrate in-phase against
U or Pu atoms.

There is a remarkably dispersive mode LO1 which has a very large group
velocity along $(001)$ direction in both materials. This mode is primarily
oxygen based and at the BZ center it is described by polarization vectors as
shown on Fig.1. We see that our theory correctly predicts a softening of
this mode as it approaches the zone boundary $X$ point although it
underestimates the experimental value of the frequency by a factor of two
for UO$_{2}$ and, most likely, also for PuO$_{2}$. Note that similar
discrepancies have been detected by us earlier in studying lattice
vibrations of Plutonium metal \cite{20}\cite{21}.

As we have gained access to the phonon dispersions, we are ready to discuss
the lattice thermal conductivity in these systems which can be expressed for
a cubic solid via the phonon frequencies $\omega_{\mathbf{q}j}$, group
velocities $v_{\mathbf{q}j}$, phonon mean free paths $l_{\mathbf{q}j}$ and
the Bose--Einstein distribution functions $N_{\mathbf{q}j}$ as follows \cite%
{1}:%
\begin{equation}
\kappa=\sum_{\mathbf{q},j}\omega_{\mathbf{q}j}v_{\mathbf{q}j}l_{\mathbf{q}j}%
\frac{\partial N_{\mathbf{q}j}}{\partial T}  \label{kappa}
\end{equation}

It is generally known that various scattering processes determine the
temperature dependent phonon mean free paths. These, for example, include
normal phonon-phonon interactions where one phonon with wave vector q gets
scattered with the creation of two phonons with wave vectors $\mathbf{q}_{1}$
and $\mathbf{q}_{2}$ such that $\mathbf{q}=\mathbf{q}_{1}+\mathbf{q}_{2}$,
and Umklapp phonon scattering processes where the quasi-momentum
conservation involves reciprocal lattice vector $\mathbf{G}$ such that $%
\mathbf{q}=\mathbf{q}_{1}+\mathbf{q}_{2}+\mathbf{G}$. At high temperatures
which we are interested in the present work, only Umklapp processes are
significant \cite{1}, and the phonon mean free paths can be evaluated from
the knowledge of the third--order anharmonicity coefficients, which can be
approximated by the dimensionless Gr\"{u}neisen constants describing the
change in the phonon frequency with respect to the atomic volume, i.e. $%
\gamma_{\mathbf{q}j}=d\ln\omega_{\mathbf{q}j}/d\ln V$. Hence, the efficiency
of each phonon mode in the heat conduction is directly related to its group
velocity but reversely proportional to the square of its Gr\"{u}neisen
constant \cite{1}.

Looking at the phonon spectra presented in Fig.2 and Fig.3, it is now clear
that the best heat carriers are either the acoustic branches or the
longitudinal optical branch LO1 which, in fact, has anomalously large group
velocity along $(001)$ (experimentally, its $v_{\mathbf{q}j}$ is twice the
LA branch!). One may then pose a question that why UO$_{2}$ or PuO$_{2}$ are
known to be such inefficient thermal conductors. When compared to most
semiconducting solids where optical modes exhibit rather weak wave vector
dispersions and thus do not participate in the heat transfer, here the
situation is much more favorable. The answer lies in the anomalously large
anharmonicity associated with the LO1 mode, making its mean free path
significantly shorter.

To make a comparative analysis, we have calculated the Gr\"{u}neisen
constants $\gamma_{\mathbf{q}j}$ associated with each vibrational mode at
the $\Gamma$ and $X$ points of the Brillouin Zone. The results of these
calculations are summarized in Table 1. As one can see the transverse
acoustic modes have relatively large anharmonicity characterized by $%
\gamma=-1.41$ and $\gamma=-1.59$ for UO$_{2}$ and PuO$_{2}$ respectively.
These values remarkably decrease for the LA modes ( $\gamma=-0.50$ and $%
-0.54 $) which at the same time show large group velocity. However, they
become huge for the most dispersive LO1 mode. Here, the frequency at $\Gamma$
shows positive $\gamma$ (equal to $0.43$ and $0.27$) meaning that it
decreases upon compression while the frequency at $X$ shows negative $\gamma$
(equal to $-2.17$ and $-2.49$) meaning its increase upon compression.
Effectively, these two effects will add up to each other and result in the
effective Gr\"{u}neisen constants equal to $-2.60$ for UO$_{2}$ and $-2.76$
for PuO$_{2}$.

\begin{table}[tbp]
\caption{Calculated phonon frequencies (in THz) and Gr\"{u}neisen constants
at $\Gamma$ and $X$ points of the Brillouin zone, the phonon group
velocities $(\times10^{2}m/s)$ estimated along $(001)$ direction as well as
contributions to lattice thermal conductivity $(Wm^{-1}K^{-1})$ at $T=1000K$
for UO$_{2}$ and PuO$_{2}$.}
\label{table}\centering
\begin{tabular}{ccccccc}
\hline
\multicolumn{7}{c}{\textbf{UO}$_{2}$} \\ \hline
\textbf{Branch} & $\omega(\Gamma)$ & $\gamma(\Gamma)$ & $\omega(X)$ & $%
\gamma(X)$ & $v_{g}$ & $\kappa$ \\ \hline
\multicolumn{1}{l}{TA} & \multicolumn{1}{r}{$0.00$} & \multicolumn{1}{r}{$%
0.00$} & \multicolumn{1}{r}{$3.88$} & \multicolumn{1}{r}{$-1.41$} & 
\multicolumn{1}{r}{$6.76$} & \multicolumn{1}{r}{$0.08$} \\ 
\multicolumn{1}{l}{LA} & \multicolumn{1}{r}{$0.00$} & \multicolumn{1}{r}{$%
0.00$} & \multicolumn{1}{r}{$6.45$} & \multicolumn{1}{r}{$-0.50$} & 
\multicolumn{1}{r}{$11.23$} & \multicolumn{1}{r}{$1.68$} \\ 
\multicolumn{1}{l}{TO1} & \multicolumn{1}{r}{$9.70$} & \multicolumn{1}{r}{$%
-1.32$} & \multicolumn{1}{r}{$9.38$} & \multicolumn{1}{r}{$-0.15$} & 
\multicolumn{1}{r}{$-0.55$} & \multicolumn{1}{r}{$\sim0$} \\ 
\multicolumn{1}{l}{LO1} & \multicolumn{1}{r}{$13.51$} & \multicolumn{1}{r}{$%
0.43$} & \multicolumn{1}{r}{$8.03$} & \multicolumn{1}{r}{$-2.17$} & 
\multicolumn{1}{r}{$-9.53$} & \multicolumn{1}{r}{$0.11$} \\ 
\multicolumn{1}{l}{TO2} & \multicolumn{1}{r}{$14.22$} & \multicolumn{1}{r}{$%
-0.20$} & \multicolumn{1}{r}{$14.42$} & \multicolumn{1}{r}{$-0.62$} & 
\multicolumn{1}{r}{$0.31$} & \multicolumn{1}{r}{$\sim0$} \\ 
\multicolumn{1}{l}{LO2} & \multicolumn{1}{r}{$16.14$} & \multicolumn{1}{r}{$%
0.62$} & \multicolumn{1}{r}{$17.81$} & \multicolumn{1}{r}{$-0.45$} & 
\multicolumn{1}{r}{$2.95$} & \multicolumn{1}{r}{$\sim0$} \\ \hline\hline
\multicolumn{7}{c}{\textbf{PuO}$_{2}$} \\ \hline
\textbf{Branch} & $\omega(\Gamma)$ & $\gamma(\Gamma)$ & $\omega(X)$ & $%
\gamma(X)$ & $v_{g}$ & $\kappa$ \\ \hline
\multicolumn{1}{l}{TA} & \multicolumn{1}{r}{$0.00$} & \multicolumn{1}{r}{$%
0.00$} & \multicolumn{1}{r}{$3.89$} & \multicolumn{1}{r}{$-1.59$} & 
\multicolumn{1}{r}{$6.67$} & \multicolumn{1}{r}{$0.08$} \\ 
\multicolumn{1}{l}{LA} & \multicolumn{1}{r}{$0.00$} & \multicolumn{1}{r}{$%
0.00$} & \multicolumn{1}{r}{$6.68$} & \multicolumn{1}{r}{$-0.54$} & 
\multicolumn{1}{r}{$11.48$} & \multicolumn{1}{r}{$1.50$} \\ 
\multicolumn{1}{l}{TO1} & \multicolumn{1}{r}{$10.21$} & \multicolumn{1}{r}{$%
-1.75$} & \multicolumn{1}{r}{$11.27$} & \multicolumn{1}{r}{$-0.43$} & 
\multicolumn{1}{r}{$1.84$} & \multicolumn{1}{r}{$\sim0$} \\ 
\multicolumn{1}{l}{LO1} & \multicolumn{1}{r}{$15.03$} & \multicolumn{1}{r}{$%
0.27$} & \multicolumn{1}{r}{$8.56$} & \multicolumn{1}{r}{$-2.49$} & 
\multicolumn{1}{r}{$-11.10$} & \multicolumn{1}{r}{$0.16$} \\ 
\multicolumn{1}{l}{TO2} & \multicolumn{1}{r}{$16.07$} & \multicolumn{1}{r}{$%
-0.52$} & \multicolumn{1}{r}{$15.42$} & \multicolumn{1}{r}{$-0.95$} & 
\multicolumn{1}{r}{$-1.02$} & \multicolumn{1}{r}{$\sim0$} \\ 
\multicolumn{1}{l}{LO2} & \multicolumn{1}{r}{$17.15$} & \multicolumn{1}{r}{$%
0.54$} & \multicolumn{1}{r}{$18.43$} & \multicolumn{1}{r}{$-0.60$} & 
\multicolumn{1}{r}{$2.46$} & \multicolumn{1}{r}{$\sim0$} \\ \hline
\end{tabular}%
\end{table}

With these data in hands we are able to estimate contributions to the
lattice thermal conductivity $\kappa$ from various phonon modes. Using a
simple Debye--like linear approximation for the phonon dispersion and
including corrections found empirically for the optical branches \cite{22},
the results of such estimates at a temperature $T=1000K$ are presented in
the last column of Table 1. We can conclude that the only efficient heat
carriers in both nuclear materials are their longitudinal acoustic phonons.
The TA modes have relatively small contribution because of their smaller
group velocities and relatively large $\gamma$, while the LO1 modes also do
not contribute due to its huge Gr\"{u}neisen constants which completely
compensate the effect of its largest $v_{\mathbf{q}j}$. The other modes have
a negligible influence on $\kappa$.

The total value of $\kappa$ in our calculation is a factor of two lower than
the experimentally known thermal conductivity of UO$_{2}$ equal to $%
3.9Wm^{-1}K^{-1}$ at $T=1000K$ \cite{23}. This is naturally connected to the
approximate character of treating anharmonic effects. However, we believe
that the relative contributions from various branches are captured correctly
in our estimates, which gives a fundamentally new insight into the processes
and factors that control thermal conductivity in these materials. As a
result, we finally discuss the origins of the large anharmonicity of the
most intriguing longitudinal optical mode LO1. As shown on Fig.1, this mode
is primarily oxygen driven. Oxygens are arranged in a cubic lattice with
Uranium or Plutonium atoms occupying every other center of the cube. It is
then clear that the structure is far from close packing and the
displacements of oxygens would involve large third order anharmonic effects.
To fix this problem, one may think of mixing these materials with elements
filling in the cubic interstitials of the lattice and preventing large ionic
excursions. One example could be oxygen overdoped materials UO$_{_{2}+x}$
and PuO$_{_{2}+x}$ where the latter one has been recently studied
theoretically \cite{7}.

To summarize, using the LDA+DMFT method and linear response theory, we have
studied lattice dynamical properties of UO$_{2}$ and PuO$_{2}.$
Contributions to lattice thermal conductivity from various phonon modes were
uncovered using the calculated group velocities and the Gr\"{u}neisen
constants. It was found that the dispersive longitudinal optical modes do
not participate in the heat transfer due to their large anharmonicity.
Material design of systems with the last effect supressed would open new
possibilities to build more efficient fuels for modern nuclear industries.

The authors are indebted to M. J. Gillan, G. Kotliar, and J. Thompson for
useful conversations. The work was supported by the NSF grants No. 0608283,
0606498 and by the US\ DOE\ grant No. DE-FG52-06NA2621.

\end{document}